\documentstyle[prb,aps,epsf]{revtex}
\begin{document}
\draft

\tighten
\twocolumn[\hsize\textwidth\columnwidth\hsize\csname @twocolumnfalse\endcsname
\title{Freezing transition of the vortex liquid in anisotropic superconductors}
\author{P.S. Cornaglia and C.A.Balseiro}
\address{Centro At\'omico Bariloche and Instituto
Balseiro,}
\address{Comisi\'on Nacional de      Energ\'{\i}a At\'omica, 8400 San Carlos 
de Bariloche, Argentina}
\maketitle

\begin{abstract}
We study the solid-liquid transition of a model of pancake vortices in laminar 
superconductors using a density functional theory of freezing. The physical 
properties of the system along the melting line are discussed in detail. We 
show that there is a very good agreement with experimental data  in the shape 
and position of the first order transition in the phase diagram and in the 
magnitude and temperature dependence of the magnetic induction jump at the 
transition.
We analyze the validity of the Lindemann melting criterion  and the 
Hansen-Verlet freezing criterion.
Both criteria are shown to be good to predict the phase diagram in the region 
where a first order phase transition is experimentally observed.  
\end{abstract}

\vskip2pc] \narrowtext

\section{Introduction}

The thermodynamics and dynamical properties of vortex matter in high
temperature superconductors have received the interest of a large
experimental and theoretical community\cite{Blatter}. One of the most
salient findings, already predicted in 1985\cite{Nelson}, is the first order
melting transition of the vortex lattice. Large thermal fluctuation in a
high $T_{c}$ material induce the melting transition first observed in
transport measurements\cite{Safar} and confirmed by a number of equilibrium
properties like direct measurements of the magnetization jump \cite{Zeld} and
latent heat\cite{Schilling}.

It is now well accepted that in clean Bi$_2$Sr$_2$CaCu$_2$O$_8$ (BSCCO) and 
untwined YBa$_2$Cu$_3$O$_7$ (YBCO) samples
the melting of the vortex lattice for an external magnetic field parallel to
the $c$-axis of the samples is first order and that the superconducting
coherence is lost simultaneously in all directions at the melting
temperature. Despite of the great amount of work devoted to the subject, the
nature of the melting transition of the vortex lattice is still
controversial.

The first theories for melting of the vortex lattice were based on the
Lindemann criterion\cite{Lindemann} and describe a situation in which the
vortex lattice melts into a liquid of entangled vortex lines. However the
behavior of the entropy change at the melting transition suggest that at
the first order transition a simultaneous loss of the triangular crystal 
structure and a decoupling of the
planes take place. In this last scenario, the liquid would be a liquid of
pancake vortices. Many authors have analyzed the experimental results
obtaining good fittings with this picture. The precise nature of the
transition and the structure of the liquid phase at the melting line is
still an open problem.

More recently, some striking regularities -weak universalities- observed
in the transport properties at the melting temperature were interpreted in
terms of the Hansen-Verlet\cite{Verlet},\cite{Rojo} freezing criterion. This
criterion, that has been proved to work very well in the case of classical
liquids, states that the liquid freezes when the first peak of the structure
factor $S(k)$ reaches a critical value. However, for the case of vortex
matter, the validity of this criterion has not been analyzed.

In this context it is interesting to revisit the density functional theory
for vortices in the highly anisotropic high $T_{c}$ superconductors. This
theory describes interacting pancake vortices in layered superconductors.
The interactions are of electromagnetic nature, Josephson interactions
involving charge transfer are neglected, and consequently it is a good
starting point to describe highly anisotropic systems. However, as we
discuss below, it may be also appropriate to describe even the
less anisotropic high $T_{c}$ cuprates. The theory allows to calculate the
melting temperature, thermodynamic properties at the melting line as well as
the validity of the phenomenological criteria for melting and freezing.

In what follows we present the model for the inter-vortex interaction and
the method for the calculus of the melting line. We present the results for
the phase diagram and compare them with experimental data. We analyze the
validity of the Lindemann and Hansen and Verlet criteria for melting and
freezing.

\section{Description of the Model}

Our starting point is the functional density theory for pancake vortices in
a layered superconductor in a magnetic field perpendicular to the layers. In
a layered system in the limit of infinite effective mass perpendicular to
the layers (zero Josephson coupling) the pancake vortices can be treated as
point like classical particles restricted to move in the planes. The
vortex-vortex interaction is given by a three-dimensional (3D) anisotropic
pair potential\cite{Feig} which in Fourier space is given by:

\begin{equation}
\beta V({\bf k})=\frac{{\Gamma \lambda }^{{2}}{[k}_{\perp }^{{2}}{+(4/d}^{{2}%
}{)sin}^{{2}}{(k}_{{z}}{d/2)]}}{{k}_{{\perp }}^{{2}}{[1+\lambda }^{{2}}{k}_{{%
\perp }}^{{2}}{+4(\lambda }^{{2}}{/d}^{{2}}{)sin}^{{2}}{(k}_{{z}}{d/2)]}},
\label{interk}
\end{equation}
here, ${\bf k}=({\bf {k}_{\perp }},{k}_{z})$ is the wave vector, $\lambda $ is
the planar London penetration depth and ${\Gamma =\beta d\Phi }_{{0}}^{{2}}{%
/4\pi \lambda }^{{2}}$ is a dimensionless strength parameter where ${\Phi }_{%
{0}}$ is the flux quantum{, }${d}${\ is the distance between planes and }$%
\beta =(k_{B}T)^{-1}$ is the inverse temperature.

Following the pioneering work of Ramakrishnan and Yussouff (RY) \cite
{Krishna1} and the extensions made to describe the vortex matter \cite
{Krishna3},\cite{Krishna2}, the functional describing the free energy
difference between the solid and liquid phases is given by: 
\begin{eqnarray}
\frac{\Delta \Omega }{k_{B}T} &=&\sum_{n}\int {d^{2}r_{\perp }}\left\{ {\rho
_{n}({\bf r}_{\perp }{\,})}\left[ {\ln \left( \frac{\rho _{n}({\bf r}_{\perp
}{\,})}{\rho _{\ell }}\right) -1}\right] {+\rho _{\ell }}\newline
\right\}  \nonumber \\
&&-\frac{1}{2}\sum_{n,n^{\prime }}\int {d^{2}r_{\perp }\,d^{2}r_{\perp
}^{\prime }\,c}_{\left| n-n^{\prime }\right| }{(|{\bf r}_{\perp }^{\prime }\,%
}{-{\bf r}_{\perp }\,|)\times }  \nonumber \\
&&{\left[ \rho _{n}({\bf r}_{\perp })-\rho _{\ell }\right] \left[ \rho _{%
{n}^{\prime }}({{\bf r}_{\perp }^{\prime }})-\rho _{\ell }\right] }.
\label{fren}
\end{eqnarray}
here, ${\bf r}$ $\equiv ({\bf r}_{\perp },nd)$ where ${\bf r}_{\perp }$ is
the in-plane coordinate and $n$ is an integer (the plane index). The quantity 
$\rho _{n}({\bf r%
}_{\perp })=<\sum_{i}\delta ({\bf r}_{\perp }-{\bf r}_{\perp i}^{n})>$ is
the vortex density with ${\bf r}_{\perp i}^{n}$ the 2D coordinate of the
i-th. particle at plane $n$, $\rho _{\ell }$ is the mean areal density of
the liquid and ${c}_{\left| n-n^{\prime }\right| }{(|{\bf r}_{\perp
}^{\prime }\,-{\bf r}_{\perp }\,|)}$ is the direct correlation function in
the liquid phase. The first integral in the right hand side of Eq. \ref{fren}
describes the non-interacting contribution to the free energy, the second
integral incorporates the effect of the interactions up to second order in
the difference $({\rho _{{\sl n}}({{\bf r}_{\perp }})-\rho _{\ell })}$.To 
evaluate this energy difference it is necessary to calculate the direct
correlation function of the liquid defined below. 

The pair distribution function is defined as:

\begin{equation}
g({\bf r},{\bf r}^{\prime })=\frac{\rho _{n,n^{\prime }}({\bf r}_{\perp },%
{\bf r}_{\perp }^{\prime })}{\rho _{n}({\bf r}_{\perp })\rho _{{\sl n}%
^{\prime }}({\bf r}_{\perp }^{\prime })}.
\end{equation}
where: 
\begin{equation}
\rho _{n,n^{\prime }}({\bf r}_{\perp },{\bf r}_{\perp }^{\prime
})=<\sum_{i\neq j}\delta ({\bf r}_{\perp }-{\bf r}_{\perp i}^{n})\delta (%
{\bf r}_{\perp }^{\prime }-{\bf r}_{\perp j}^{\,n^{\prime }})>
\end{equation}
here $\rho _{n,n^{\prime }}({\bf r}_{\perp },{\bf r}_{\perp }^{\prime })$ is
the probability density of finding two particles at ${\bf r=}({\bf r}_{\perp
},nd)$ and ${\bf r}^{\prime }{\bf =}({\bf r}_{\perp }^{\prime },n^{\prime
}d) $ respectively. Due to the symmetry of the liquid $g({\bf r},{\bf r}%
^{\prime })=g_{\left| n-n^{\prime }\right| }(\left| {\bf r}_{\perp }-{\bf r}%
_{\perp }^{\prime }\right| ).$

The direct correlation function $c_{n}({\bf r}_{\perp })$ is defined by
means of Ornstein-Zernike equation \cite{ozie}: 
\begin{equation}
h_{n}(r_{\perp })=c_{n}(r_{\perp })+\rho _{\ell }\sum_{n^{\prime }}\int
d^{2}r_{\perp }^{\prime }c_{\left| n-n^{\prime }\right| }(|r_{\perp
}-r_{\perp }^{\prime }|)h_{n^{\prime }}(r_{\perp }^{\prime }).  \label{OZ}
\end{equation}
where $h_{n}(r_{\perp })=g_{n}(r_{\perp })-1$ is the pair correlation
function$.$

This integral equation for the direct correlation function $c_{n}(r_{\perp
}) $ can be solved in the hipernetted chain approximation, and inserting the
solution in Eq.(\ref{fren}), the free energy difference between the solid
and liquid phase is given only as a functional of the density $\rho _{n}(%
{\bf r}_{\perp }{\,})$. At this point, we can expand the density in a set of
trial reciprocal-lattice vectors:

\begin{equation}\label{dens}
\rho _{n}({\bf r}_{\perp }{\,})/\rho _{\ell}=1+\eta +\sum_{{\bf G\neq 0}}\rho 
_{%
{\bf G}}{{e}^{i{\bf G\cdot r}}}
\end{equation}

The quantities $\eta $ and $\rho _{{\bf G}}$, may be viewed as dimensionless
order parameters, $\eta $ is the fractional density difference between the
liquid and the solid phases and the $\rho _{{\bf G}}$ are the Fourier components
of the density in the solid phase. Minimizing the free energy (\ref{fren})
with respect to these parameters, we obtain the following equation:

\begin{equation}
1+\eta +\sum_{{\bf G\neq 0}}\rho _{{\bf G}}{e}^{i{\bf G\cdot r}}=\exp ({\rho
_{\ell }c_{0}\eta +\rho _{\ell }\sum_{{\bf G\neq 0}}\rho _{{\bf G}}c_{{\bf G}%
}{e}^{i{\bf G\cdot r}}}),  \label{auto}
\end{equation}
where $c_{{\bf G}}$ is the Fourier transform of the direct correlation
function and $c_{0}\equiv c_{{\bf G=0}}$.

Due to the small compressibility of the system the fractional density change
is small ($\eta \ll 1$), and as a first approximation can be neglected in
the first term of (\ref{auto}). Note that the product ${c_{0}\eta }$ is not
necessarily small and the first term in the exponential of the right hand
side of (\ref{auto}) has to be retained. Transforming Fourier Eq. (\ref{auto}) 
we obtain after some algebra a set of coupled equations for the $\rho _{
{\bf G}}$:

\begin{equation}
\rho _{{\bf G}}=\frac{\sum_{n}\int_{A_{c}}d^{2}r_{\perp }e^{-i{\bf G\cdot r}
}\exp ({\rho _{\ell }\sum_{{\bf G\neq 0}}\rho _{{\bf G}}c_{{\bf G}}{e}^{i
{\bf G\cdot r}})}}{\sum_{n}\int_{A_{c}}d^{2}r_{\perp }\exp ({\rho _{\ell
}\sum_{{\bf G\neq 0}}\rho _{{\bf G}}c_{{\bf G}}{e}^{i{\bf G\cdot r}})}}.
\label{equrhog}
\end{equation}
here $A_{c}$ is the area of the in-plane unit cell. Given the Fourier
components $c_{{\bf G}}$ of the direct correlation function, these equations
can be solved for \{$\rho _{{\bf G}}$\}. Before presenting the numerical
results, the following points deserve a comment:

$i)$ A uniform liquid is always a solution of (\ref{equrhog}) \{$\rho _{{\bf 
G}}=0$\}. As the $c_{{\bf G}}$ increase, non-uniform solutions may suddenly
appear (with $\rho _{{\bf G}}\sim 0.5$) and correspond to the solid phase
with a spatially periodic density.

$ii)$ At the transition the free energy difference $\Delta \Omega $ defined
in (\ref{fren}) is equal to zero. To calculate the transition line in the $
B-T$ phase diagram, we have to calculate the direct correlation function as
a function of the magnetic induction and temperature and use it as an
input in the RY density functional theory. The transition is obtained when a
set of order parameters \{$\rho _{{\bf G}}\neq 0$\} given by (\ref{equrhog})
satisfy the condition $\Delta \Omega =0$.

$iii)$ As discussed in previous works, the numerical calculation can be
simplified by noting that in the liquid phase, the Fourier transform $c_{
{\bf G}}$ of the direct correlation function rapidly decay with increasing $
\left| {\bf G}\right| $, it is then a good approximation to truncate the
series ${\sum_{{\bf G\neq 0}}\rho _{{\bf G}}c_{{\bf G}}{e}^{i{\bf G\cdot r}}}
$ in (\ref{equrhog}) and take only a few terms with $c_{{\bf G}}\neq 0$. The
number of terms retained defines the number of self consistent order
parameters \{$\rho _{{\bf G}}$\} included in the theory. In what follows we
consider only two order parameters corresponding to the first two peaks of
the in-plane direct correlation function as first suggested by Ramakrishnan
\cite{Krishnad} for a 2D system.

$iv)$ With the resulting set of order parameters that characterize the solid
phase at the melting point in the limit $\eta \ll 1$, the fractional density
change can be estimated. However in the RY approximation used here the
results for $\eta $ are not accurate, in particular it is ease to show that $
\eta >0$ in contradiction to what is expected for the case of vortices.
Rather than improving the estimate of $\eta $ by including higher order
terms in the functional $\Delta \Omega $ of Eq. (\ref{fren}) -which are
given by not available three particle correlation functions- we estimate the
entropy change at the transition and calculate $\eta $ by means of the
Clausius-Clapeyron relation. The higher order corrections to $\Delta \Omega $
are not strongly coupled to the order parameters \{$\rho _{{\bf G}}$\} and
the present approximation is known to give accurate results when the
crystal structure of the solid phase is given.

\section{Correlation functions}

In this section we present results for the pair correlations in the liquid
phase, we use the hipernetted chain approximation (HNC) which is well
described in the literature. For the sake of completeness we only give here
the basic formulas of classical liquids theory relevant for our case.

The pair correlation function can be written in the form:

\begin{equation}
g_{n}(r_{\perp })={e}^{[g_{n}(r_{\perp })-1-\beta V_{n}(r_{\perp
})-c_{n}(r_{\perp })+B_{n}(r_{\perp })]},  \label{clausu}
\end{equation}
where $B_{n}(r_{\perp })$ is known as the bridge function and its closed
expression is not known for an arbitrary potential. In the HNC approximation
we have $B_{n}(r_{\perp })\equiv 0$ which is an excellent approximation for
long ranged soft potentials as the one we are dealing with \cite{ozie}. In the
HNC approximation Eq. (\ref{clausu}) can be rewritten in the following form:

\begin{equation}
c_{n}(r_{\perp })={e}^{[-\beta V_{n}(r_{\perp })+Y_{n}(r_{\perp
})-1-Y_{n}(r_{\perp })]},  \label{HNC}
\end{equation}
where $Y_n(r_\perp)=h_n(r_\perp)-c_n(r_\perp)$ . This equation together with 
the Ornstein Zernike
relation (Eq.(\ref{OZ})) are the basic equations that allows to evaluate the
pair correlation functions. The asymptotic behavior of the direct
correlation function as $r_{\perp }\to \infty $ and $n\to \infty $ is known
to be $c_{n}(r_{\perp })=-V_{n}(r_{\perp })/k_{B}T$. For the
numerical calculation it is then convenient to divide the correlation
functions in short and long ranged contributions according to:

\begin{equation}
c_{n}(r_{\perp })=c_{nS}(r_{\perp })+c_{nL}(r_{\perp }),
\end{equation}
with 
\begin{equation}
c_{nL}(r_{\perp })=-\beta V_{n}(r_{\perp })
\end{equation}
and 
\begin{equation}
Y_{n}(r_{\perp })=Y_{nS}(r_{\perp })+Y_{nL}(r_{\perp }),
\end{equation}
with 
\begin{equation}
Y_{nL}(r_{\perp })=+\beta V_{n}(r_{\perp }).
\end{equation}

Due to the highly anisotropic nature of the pair potential, as a first
approximation we take the short ranged contribution different from zero only
for the in-plane correlations: 
\begin{equation}
c_{nS}(r_{\perp })={\delta }_{n,0}c_{0S}(r_{\perp })
\end{equation}
and the problem reduces to calculate a single function ($c_{0S}(r_{\perp })$
) given by the following set of integral equations:

\begin{equation}
c_{0S}(r_{\perp })=\exp (Y_{0S}(r_{\perp }))-1-Y_{0S}(r_{\perp }),
\label{OZ_lay}
\end{equation}

\begin{equation}
Y_{0S}(k_{\perp })=\frac{d}{2\pi }\int_{-\frac{\pi }{d}}^{\frac{\pi }{d}
}dk_{z}\frac{c_{0S}(k_{\perp })-\beta V({\bf k})}{1-\rho_{\ell}[c_{0S}(k_{\perp
})-\beta V({\bf k})]}-c_{0S}(k_{\perp }),  \label{HNC_lay}
\end{equation}
the last equation is the Fourier transformed Ornztein-Zernike equation. This
expression can be further simplified by an analytic integration due to the
particular form of the pair potential. 

Finally the out of plane short ranged
correlation functions in a second order approximation are given by the
following equation:

\begin{equation}
Y_{nS}(k_{\perp })=\frac{d}{2\pi }\int_{-\frac{\pi }{d}}^{\frac{\pi }{d}
}dk_{z}{e}^{-ik_{z}nd}\frac{c_{0S}(k_{\perp })-\beta V({\bf k})}{
1-\rho_{\ell}[c_{0S}(k_{\perp })-\beta V({\bf k})]}.  \label{HNC_layn}
\end{equation}

In Fig. \ref{paird} we present results for the pair distribution
function where the lengths are in units of the characteristic distance $a$ 
given by $\pi a^2 \rho_\ell =1$. The first peak of the in plane pair 
distribution function occurs
at a finite value of the $r_{\perp }$which is given by the mean
inter-particle distance $(\sim 2a)$. For the out of plane distribution function 
the
maximum occurs at $r_{\perp }=0$ which shows the tendency of pancake
vortices to form stacks. In Fig. \ref{decay} the behavior of $h(r_{\perp
}=0,n\neq 0)$ is presented and shows an exponential decay as a function of $
n $. This behavior allows as to define a correlation length $\Lambda $
through the following relation: $h(r_{\perp }=0,n)\propto \exp (-dn/\Lambda
).$ In Fig. \ref{lambda} we present the behavior of $\Lambda $ as a function
of the magnetic induction $B$, that can be fitted with $\Lambda \propto 1/
\sqrt{B}$. Other authors \cite{Krishna3} had an excellent fit of their data
in a smaller range of fields with a function of the form $\Lambda
=A_{1}+A_{2}/B.$

It is important to note that the characteristic length $\Lambda $ describes
the long distance behavior of the correlation functions. The short distance
behavior is given by $h(r_{\perp }=0,n)$ with small $n$ which in general
decays very fast. As can be seen in Fig.\ref{decay}, $h(r_{\perp }=0,n=1)$ is 
very
small ($\simeq 10^{-2}$), and although its value depends on field and
temperature, in the liquid phase, it always remains much smaller than $1$.

As the temperature decreases the system develops stronger correlations. This
is clearly observed in the structure factor, defined by:

\begin{equation}
S({\bf k}_{\perp },k_{z})=1+\rho _{\ell }\sum_{n}\int d^{2}r_{\perp
}\,h_{n}(r_{\perp })\,{e}^{i({\bf k}_{\perp }\cdot {\bf r}_{\perp
}+k_{z}nd)}.
\end{equation}

In Fig. \ref{strfac} the structure factor is shown for two different
temperatures, the lower the temperature, the higher the first peak in $S(
{\bf k})$.

In the next section we use these results to calculate the melting line.

\section{Phase diagram}

We used the correlation functions calculated above as an input to the RY
density-functional theory of freezing. In Fig. \ref{pdiag1} we show the
results obtained for the phase diagram with parameters corresponding to
BSCCO but without temperature dependence: $\lambda(T)=\lambda(0)=1500$\AA , 
$d=15$\AA $\,$ and $T_{c}=90K$. To
understand these results we resort to a simple model obtained by mapping the
vortex system onto a 2-D boson system. In this model, the free energy of a
system of vortices is given by\cite{Nelsone}:

\begin{equation}
{\rm F}[r(z)]=\int_{0}^{L}\,dz\left\{ \sum_{i}\frac{1}{2}g\left( \frac{
dr_{i}(z)}{dz}\right) ^{2}+\sum_{i<j}V(r_{ij}(z))\right\} ,
\end{equation}
where $r_{i}(z)$ determines the position and shape of the i-th. vortex, $g$
is the elastic energy and the last term describes the interaction between
vortices, which is taken to be logarithmic, $r_{ij}(z)$ is the distance
between vortices. In a solid phase, of lattice constant $a_0$, we can use an 
harmonic approximation for the pair interaction
potential. Assuming a characteristic length $l$ which defines the
correlation along the field direction, the free energy after a thermodynamic
average can be approximated by:

\begin{equation}
{{\rm F}^{\prime }}={\rm F}/N\simeq \frac{<r^{2}>}{2}\left\{ \frac{g\,}{l}
+kl\right\} ,  \label{fren3}
\end{equation}
where ${{\rm F}^{\prime }}$ is the mean free energy per characteristic vortex
segment of length $l$, $N$ is the number of independent segments in the
sample, $<r^{2}>$ is the mean square displacement of the vortices and $
k\propto \epsilon _{0}/a_{0}^{2}$ is the curvature of the logarithmic pair
potential at the mean inter-particle distance. Using the Lindemann melting
criterion $<r^{2}>=u_{L}^{2}a_{0}^{2}$, with $u_{L}\sim 0.1-0.3$ along the
transition line and the equipartition theorem that states that ${{\rm F}
^{\prime }\sim k}_{B}T$, the expression above reduces to:

\[
{k}_{B}T_{m}=\frac{u_{L}^{2}a_{0}^{2}}{2}\left( \frac{g}{l}+kl\right) . 
\]
If the characteristic length $l$ is taken as the distance between planes
which should be a good approximation at least in the high field regime we
obtain:

\begin{equation}
k_{B}T_{m}\simeq \frac{u_{L}^{2}\epsilon _{0}d}{2}\left\{ 1+\frac{g\Phi _{0}
}{d^{2}\epsilon _{0}B_{m}}\right\} .  \label{meltB}
\end{equation}

If the inter-plane correlations were relevant at the melting point, the
characteristic length $l$ should be obtained minimizing the free energy (\ref
{fren3}). In this case the melting line is given by:

\begin{equation}
k_{B}T_{m}=u_{L}^{2}\sqrt{g\epsilon _{0}}\left( \frac{\Phi _{0}}{B_{m}}
\right) ^{\frac{1}{2}}.  \label{meltA}
\end{equation}

The density functional data for the melting line were fitted with the
following expression $T_{m}=T_{m}^{2D}+A_{1}/B_{m}$ according to (\ref{meltB}). 
This type of behavior is expected for a highly anisotropic system at high
fields, when the liquid state at the melting temperature presents very short
correlations along the $c$ axis{\bf . }Although the fit is less accurate for
small fields we never obtain a behavior of the type given by (\ref{meltA}).
This suggests that in our field and temperature range, the system
simultaneously melts and the planes decouple to form a liquid of pancake
vortices. The less accurate fit for the low field data could indicate that
we are entering a cross-over to a line melting regime. In the limit of high
fields the melting temperature goes asymptotically to the 2D melting
temperature $T_{m\text{ }}^{2D}$. This is due to the fact that at high
fields the inter-plane correlations become irrelevant and the system behaves
as a collection of independent planes. The 2D melting temperature, obtained
with a functional density theory in the RY approximation, is lower than
expected from a Kosterlitz-Thouless dislocation unbinding theory.
This is presumably due to the fact that the HNC approximation underestimates
the correlations in the 2D case as was discussed by other authors\cite
{Krishna3}.

In order to compare our results with the experimental data, we included the
temperature dependence of the penetration length\cite{Leel}: $\lambda(T) 
=\lambda
(0)/[1-(T/T_{c})^{4}]^{1/2}$. In Fig. \ref{zeldov} we present our
results with $\lambda (0)=1500$\AA $\,$ together with the experimental
results of Zeldov et al. In the experimental data the first order transition
ends at a critical point presumably due to point-like pinning centers, an
effect not included in this functional density theory. In order to make this
comparison, $\lambda (0)$ was used as a fitting parameter. A more realistic
value of $\lambda (0)$ for BSCCO is $\sim 2000$\AA . If the $\lambda (0)$
is increased, the melting line moves to lower fields due to the decrease of
the intensity of the pancake-pancake interactions, this could be compensated
by the inclusion of a small Josephson coupling between planes in a more
realistic treatment.

\section{Entropy and magnetic induction jump at $T_{m}$}

The entropy jump can be calculated as the temperature derivative of the free
energy difference:

\begin{equation}
\Delta S=-\left( \frac{\partial \Delta \Omega }{\partial T}\right) _{V,H}
\label{entropia1}
\end{equation}

Using Ecs. \ref{dens} and \ref{auto} the free energy difference of 
Ec.\ref{fren} can be written as:

\begin{eqnarray}
\frac{\Delta \Omega }{k_{B}T} &=&-\ln {\left[ \frac{1}{v_{c}}{
\int_{v_{c}}d^{3}r\exp {\left( \sum_{{\bf G}}\rho _{{\bf G}}c_{
{\bf G}}e^{i{\bf G}\cdot {\bf r}}\right) }}\right] }  \nonumber \\
&&+\frac{1}{2}\sum_{{\bf G}}{c_{{\bf G}}}{\rho _{{\bf G}}^{2}}\,,
\label{difenfour}
\end{eqnarray}

To evaluate the entropy jump we must calculate the temperature derivatives
of the Fourier transforms of the liquid direct correlation functions ${c_{
{\bf G}}}$ at constant $H$ and $V$, that are not known. Following reference 
\cite{Krishnad}, this temperature derivatives can be estimated on the solid 
phase using the fluctuation-dissipation theorem
in the classical limit. This procedure gives for the structure factor $S({\bf 
G})$ a
linear dependance on temperature. Using this result we obtain the following
expression for the entropy jump:

\begin{equation}
\Delta s/k_{B}=3[\rho _{1}^{2}(1-c_{1})+\rho _{2}^{2}(1-c_{2})],
\end{equation}
in our two order parameter calculation.

In Fig. \ref{entrop} we present the density functional results for the
entropy change. Its average value $\sim 0.3k_{B}$ is consistent with
experimental data\cite{Zeld}. With the entropy jump result and the
Clausius-Clapeyron relation we can calculate the magnetic induction jump at
the transition line:

\begin{equation}
\frac{\Delta B}{B_{m}}=\left( \frac{d\Phi _{0}}{4\pi \Delta s}\frac{dB_{m}}{
dT}\right) ^{-1},
\end{equation}
where we have made the following approximation: 
\begin{equation}
\frac{dH_{m}}{dT}\sim \frac{dB_{m}}{dT}.
\end{equation}
The results for the relative magnetic induction jump are plotted in Fig. \ref
{bjump} with the Zeldov et al. experimental data digitalized. The parameters
used to obtain this data were the same as those used to calculate the phase
diagram of Fig. \ref{zeldov}. There is a very good agreement in the
magnitude and temperature dependance with this direct experimental
measurement.

\section{Melting and freezing criteria}

In this section we analyze the Lindemann melting criterion and the
Hansen-Verlet freezing criterion for the vortex system. The Lindemann
criteria states that the solid melts when the square root of the mean square
displacement of the particles is an order of magnitude smaller than the
lattice parameter i.e. $<u^{2}>/a_{0}^{2}=u_{L}^{2}$, with $u_{L}\sim 0.1$.
This criteria has been widely used to estimate the phase diagram and
different properties of the system at the melting line. In general it is
accepted that for the melting of vortex mater the Lindemann parameter $u_{L}$
is larger than for other classical 3D solids and its value is usually taken
in the range $0.2-0.3$.

We estimated the mean square displacement as:

\begin{equation}
<u^{2}>\simeq \frac{1}{A_{c}}\int_{A_{c}}d^{2}r_{\perp }\,r_{\perp }^{2}
\frac{\rho _{n}({\bf r}_{\perp }\,)}{\rho _{\ell }},
\end{equation}
the density at the melting line is obtained using (\ref{auto}) and the
results for the Lindemann parameter are shown in Fig. \ref{lindemann}. For
low fields, numerical results show that $u_{L}$ $\simeq $ $0.2$ in good
agreement with previous estimates, as the field is increased (the melting
temperature decreases) the Lindemann parameter increases. This is due to the
fact that, as discussed above, in the present model as the field increases
the interplane coupling decreases and the transition becomes essentially 2D.
For a 2D transition, the Lindemann parameter is larger than for a 3D one.

The Hansen-Verlet freezing criteria is the counterpart of the Lindemann
criteria and states that a liquid freezes when the first peak of the
structure factor reaches the value $2.85$ in a 3D classical system and of
the order of $5.5$ for a 2D plasma \cite{plasma2d}. Having evaluated the
structure factor at the melting transition within the functional density
theory, we can show the validity of this criteria for the case of freezing
of the vortex liquid. The results for $S(k_{\max })$ are shown in Fig. \ref
{lindemann} and as in the case of the Lindemann criteria the Hansen-Verlet
criteria is valid for low fields with a value $S(k_{\max })\sim 6.5$, much
larger than that obtained for isotropic 3D system. As the melting
temperature decreases $S(k_{\max })$ towards its two dimensional value. In
our case the value of $S(k_{\max })$ at small fields is larger than the 2D
value due to the following reason: the idea behind the Hansen-Verlet
criteria is that when the liquid develops correlations, that overcame some
critical value, it freezes. At high temperatures the first peak in $S(k)$
dominates and is a good measure of the correlations, as the temperature
decreases, the liquid develops stronger correlations for large ${\bf G}$ and
both the first and second peaks in $S(k)$ became relevant at the freezing
point. Consequently, it is not necessary for $S(k_{\max })$ to become too
large to indicate that the liquid has developed critical correlations. This
can be viewed as a relative increase of the order parameter $\rho _{2}$ with
respect to $\rho _{1}$at the transition as the melting temperature
decreases. The results indicate that for low fields the ratio $\rho
_{1}/\rho _{2}$ is larger than $4$, while at high fields is of the order of $
2$.

\section{Summary and discussion}

In this work we study the melting transition for the vortex system in
layered superconductors using the RY functional density theory. Using
reasonable set of parameters for BSCCO we obtained qualitative and
quantitative agreement with the available experimental data.

The functional density theory allows an estimation of the entropy jump at
the transition and using the Clausius-Clapeyron relation we estimated the
jump in magnetic induction. The same parameters used to fit the phase
diagram, give results for the magnetic induction jump that are in
quantitative agreement with the experimental results. Finally we analyzed
the validity of the melting and freezing criteria and showed that in the
region of interest -the low field region where the first order transitions
occurs- the Lindemann and the Hansen-Verlet criteria are valid within a
error of $10\%$ which is typical for this type of criteria.

The model predicts the melting of the vortex lattice into a liquid of almost
decoupled pancake vortices even in the low field region. In this sense we
may interpret our results as a simultaneous decoupling and melting of the
vortex lattice\cite{Glazman} rather than a melting into a liquid of vortex
lines. This picture may be appropriate for a highly anisotropic system like
BSCCO. However, in a recent paper Kitazawa et al.\cite{Kitazawa}
successfully used a scaling suitable for simultaneous decoupling and melting
not only for BSCCO but also for less anisotropic systems like YBCO. This
scaling predicts that the melting line is given by:

\begin{equation}
B_{m}=C\left( \frac{T_{c}}{T_{m}}-1\right)  \label{scale}
\end{equation}
where the prefactor depends on the anisotropy and interplane separation of
the system. Both the experimental results of Zeldov et al. for BSCCO and our
numerical results for low fields are in good agreement with this expression
as shown in Fig. \ref{sublim}. In the figure we present results corresponding
to a zero temperature penetration depth $\lambda (0)=1500$\AA\ which gives
very good agreement with the results of Zeldov et al. for the low field
region where the first order transition is experimentally observed. As
discussed in section IV, a more realistic value for $\lambda (0)$ is $2000$
\AA , which gives a melting line laying below the experimental points.
However, in any real system there is some Josephson coupling between the
planes that should compensate the lowering of the electromagnetic coupling
due to the increase of $\lambda (0)$. In this sense, we take $\lambda (0)$
as an effective parameter that measures the strength of the coupling between
planes. A change in $\lambda (0)$ does not change the temperature dependance
of $B_{m}$ for low fields, but only the coefficient $C$ in Eq.(\ref{scale}).

If experimental results confirm the fact that even for less anisotropic
system, the first order melting transition correspond to a simultaneous
decoupling and melting of the vortex lattice, the present theory could be
used to describe these systems provided that $\lambda (0)$ or the interplane
distance are chosen as effective parameters. In particular the validity of
the melting and freezing criteria obtained for the highly anisotropic
laminar BSCCO could be extended to other less anisotropic systems like YBCO.

Work partially supported by grants CONICET $4946/96$ and ANPCYT PICT $0251/98$. 
One of us (C.A.B.) is partially supported by CONICET.

\begin{figure}
\epsfysize = 6cm
\begin{center}
\leavevmode
\epsffile{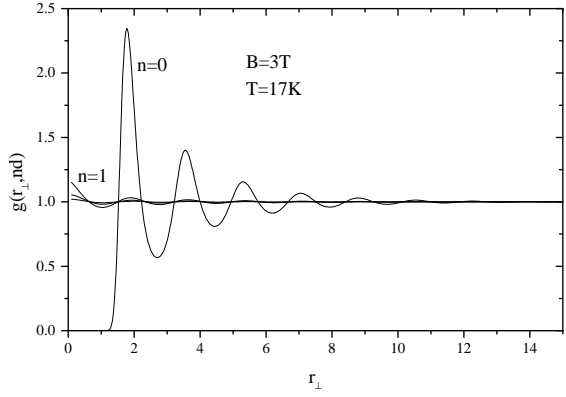}
\end{center}
\caption{HNC results for the pair distribution function $g(r_
\perp,n=0,1,2,\ldots)$ of the vortex liquid for $B=3T$ and $T=17K$. For $n=0$,
the first maxima as a function of $r_\perp$ is given by the mean particle 
spacing
and for $n\neq 0$ is at $r_\perp=0$.}
\label{paird}
\end{figure}

\begin{figure}
\epsfysize = 6cm
\begin{center}
\leavevmode
\epsffile{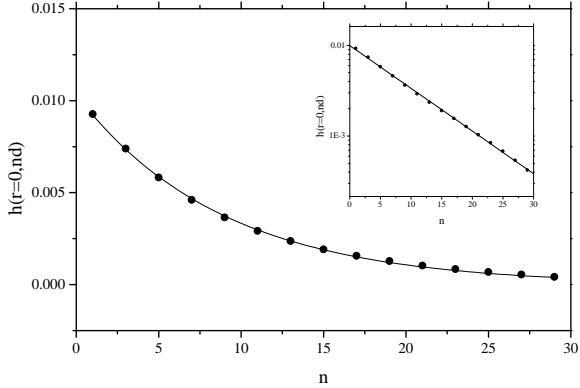}
\end{center}
\caption{The pair correlation $h(r_\perp,n)$ at $r_\perp=0$ as a function of 
$n$ for $B=0.065T$ and $T=31K$. The values can be fitted with an exponential
decay, as shown in the inset with a log-normal scale.}
\label{decay}
\end{figure}

\begin{figure}
\epsfysize = 6cm
\begin{center}
\leavevmode
\epsffile{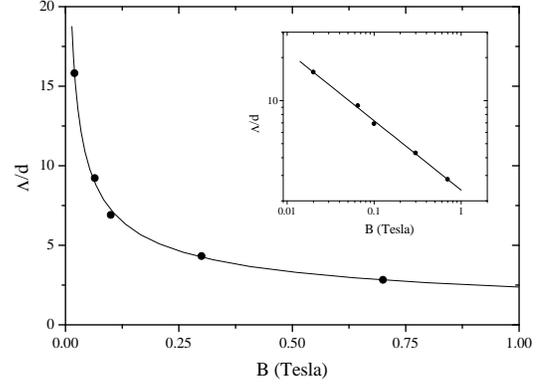}
\end{center}
\caption{The c-axis correlation length as a function of the magnetic induction 
$B$ with $T=31K$. The slope of the linear fit in the inset is 1/2.}
\label{lambda}
\end{figure}

\begin{figure}
\epsfysize = 6cm
\begin{center}
\leavevmode
\epsffile{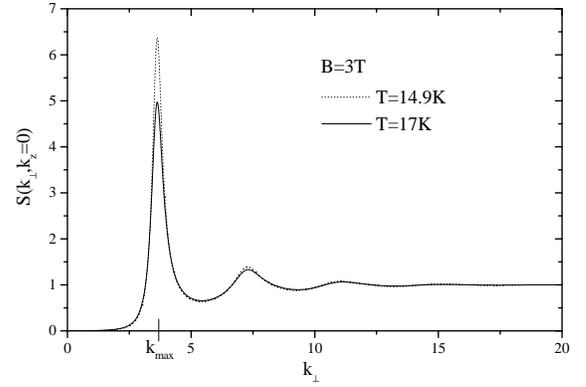}
\end{center}
\caption{Structure factor for $k_z=0$. Note the
increase in the structure as the temperature is lowered. The amplitude of the
first peak is the quantity that we considered for the Hansen-Verlet
criterion. }
\label{strfac}
\end{figure}

\begin{figure}
\epsfysize = 6cm
\begin{center}
\leavevmode
\epsffile{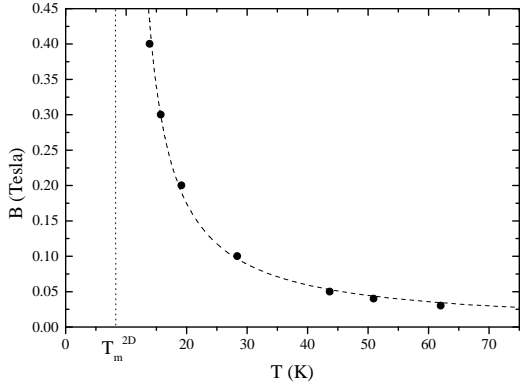}
\end{center}
\caption{Liquid-Solid transition line in the B-T phase diagram for the
vortex system in BSCCO. The solid line is fit with $B=A_1/(T-T_m^{2D})$,
where $A_1$ and $T_m^{2D}$ are variational parameters.}
\label{pdiag1}
\end{figure}

\begin{figure}
\epsfysize = 6cm
\begin{center}
\leavevmode
\epsffile{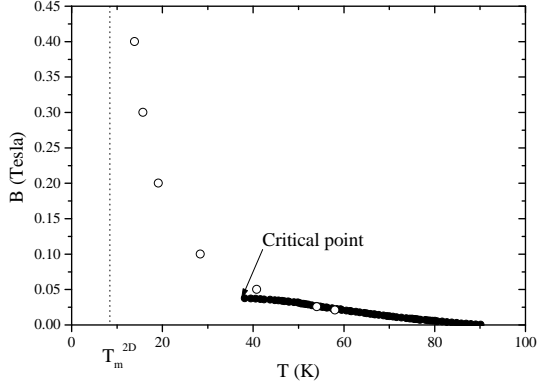}
\end{center}
\caption{Phase diagram of BSCCO. The open dots are the RY functional
density results and the solid dots are direct meassurements by Zeldov
et al. In the experimental data there is a critical point where the first
orther transition ceases to exist, which is pressumably caused by the presence 
point
pinning centers in the sample. }
\label{zeldov}
\end{figure}

\begin{figure}
\epsfysize = 6cm
\begin{center}
\leavevmode
\epsffile{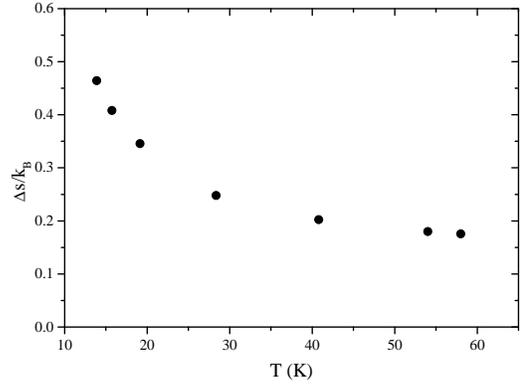}
\end{center}
\caption{Entropy jump at the first order transition calculated with the RY
functional density of freezing.}
\label{entrop}
\end{figure}

\begin{figure}
\epsfysize = 6cm
\begin{center}
\leavevmode
\epsffile{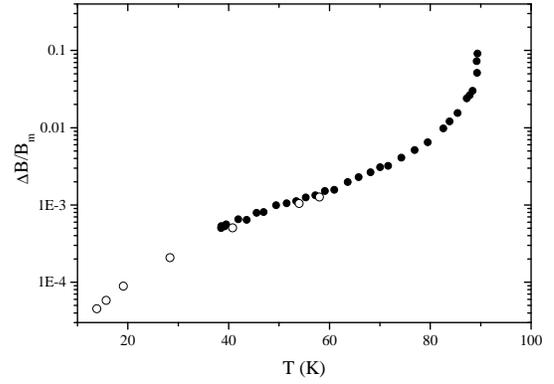}
\end{center}
\caption{Relative magnetic induction jump at the first order transition.
The open dots were calculated using the Clausius-Clapeyron relation and the RY 
functional
density theory of freezing. It is in very good agreement with the
experimental results of Zeldov et al. (solid dots) in the magnitude and in the
temperature dependence.}
\label{bjump}
\end{figure}

\begin{figure}
\epsfysize = 6cm
\begin{center}
\leavevmode
\epsffile{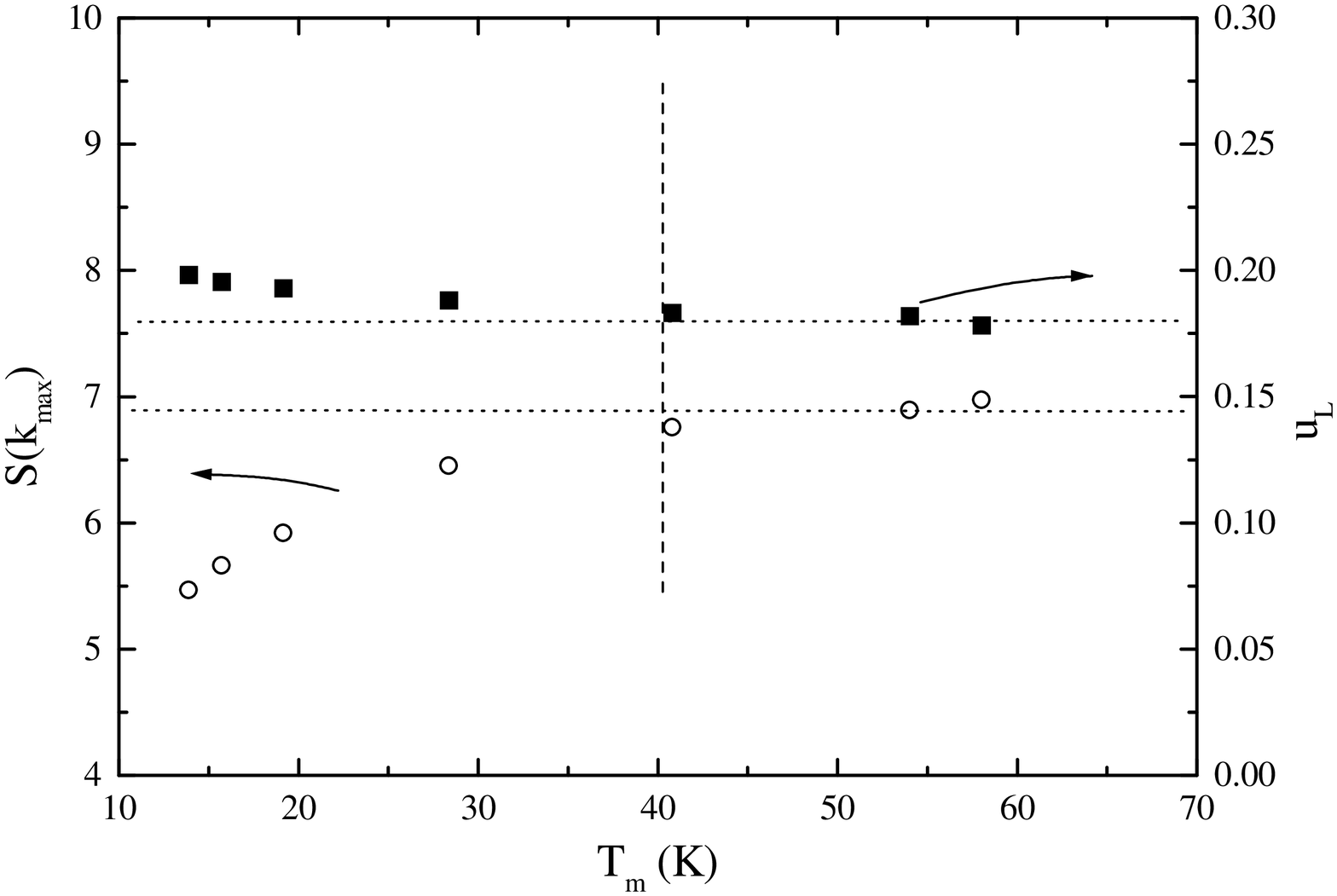}
\end{center}
\caption{Lindemann melting parameter and Hansen-Verlet freezing parameter as
a function of $T_m$ for BSCCO calculated in the RY approximation. Both
parameters are approximatly constant in the region where the first order phase 
transition it is experimentally observed. }
\label{lindemann}
\end{figure}

\begin{figure}
\epsfysize = 6cm
\begin{center}
\leavevmode
\epsffile{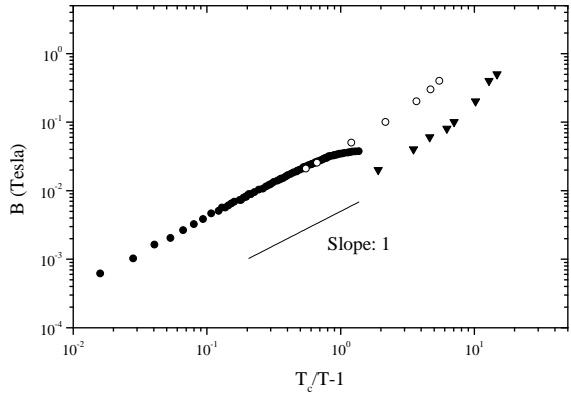}
\end{center}
\caption{$B_{m}$ as a function of $T_{c}/T-1$. A test for sublimation.
Calculations with different $\lambda (0)$ or $d$ give liquid-solid
transition lines displaced paralel to each other giving the same effect of a
different anysotropy.}
\label{sublim}
\end{figure}

\end{document}